\documentclass[twocolumn,showpacs,preprintnumbers,amsmath,amssymb]{revtex4}
\usepackage{graphicx}% Include figure files
\usepackage{dcolumn}% Align table columns on decimal point
\usepackage{bm}% bold math

\begin{document}

%\preprint{APS/123-QED}

\title{Minimal Models for a Superconductor-Insulator Conformal Quantum Phase Transition}% Force line breaks with \\

\author{M. Cristina Diamantini}
\altaffiliation[On leave of absence from: ]{INFN and Dipartimento di Fisica, University of Perugia, via A. Pascoli, I-06100 Perugia, Italy}%Lines break automatically or can be forced with \\
\email{cristina.diamantini@pg.infn.it}
\affiliation{%
Theory Division, CERN, CH-1211 Geneva 23, Switzerland 
}%

%\author{Pasquale Sodano}
%\altaffiliation[Also at ]{Perimeter Institute of Theoretical Physics 31, Caroline St. North, Waterloo, Ontario N2L2Y5, Canada}%Lines break automatically or can be forced with \\
%\email{pasquale.sodano@pg.infn.it}
%\affiliation{%
%INFN and Dipartimento di Fisica, University of Perugia, via A. Pascoli, I-06100 Perugia, Italy
%}%

\author{Carlo A. Trugenberger}
%\altaffiliation[Also at ]{Theory Group, Physics Department, CERN, CH-1211 Geneva 23, Switzerland}%Lines break automatically or can be forced with \\
\email{ca.trugenberger@bluewin.ch}
\affiliation{%
SwissScientific, chemin Diodati 10, CH-1223 Cologny, Switzerland
}%

%\author{M. Cristina Diamantini}
%\affiliation{

\date{\today}% It is always \today, today,

\begin{abstract}
Conformal field theories do not only classify 2D classical critical behavior but they also govern a certain class of 2D quantum critical behavior. In this latter case it is the ground state wave functional of the quantum theory that is conformally invariant, rather than the classical action. We show that the superconducting-insulating (SI) quantum phase transition in 2D Josephson junction arrays (JJAs) is a (doubled) $c=1$ Gaussian conformal quantum critical point. The quantum action describing this system is a doubled Maxwell-Chern-Simons model in the strong coupling limit. We also argue that the SI quantum transitions in frustrated JJAs realize the other possible universality classes of conformal quantum critical behavior, corresponding to the unitary minimal models at central charge $c=1-6/m(m+1)$.

\end{abstract}

\maketitle

%\section{Introduction}

Conformal field theories describe the critical behaviour of systems at second order phase transitions \cite{difra}. In three dimensions, conformal invariance does not provide much more information than simple scale invariance. In two dimensions, instead, the infinite dimensionality of the conformal algebra leads to significant restrictions on allowed conformal field theories and, ultimately to a complete classification of possible 2D critical phenomena.  

While this program has been fully carried out, recent research has focused on quantum phase transitions, which are driven by quantum, rather than thermal fluctuations and which are characterized by quantum critical couplings rather than critical temperatures \cite{sen}. Such quantum phase transitions are thought to separate topological states of matter and are often characterized by emergent gauge fields and fractionalization. 

In a seminal contribution \cite{fra} it was shown that there exist 2D second-order quantum critical points that are also described by conformal field theories. In this case it is not the action of the effective field theory at the critical point that is conformally invariant but rather the ground state wave functional of the quantum theory. The idea is as follows. Suppose that the Schr\"odinger picture ground state wave functional of a quantum theory at a critical point is given by
\begin{equation}
\Psi [ \phi ] = {\rm e}^{-{\lambda \over 2} \int d^2 {\bf x} \ \partial_i \phi \partial_i \phi} \ . 
\label{one}
\end{equation}
Then,  the equal-time quantum correlations of the dynamical fields $\phi$ are given by 
\begin{eqnarray}
&&\langle \phi({\bf x}_1) \dots \phi({\bf x}_n) \rangle = {1\over Z} \int {\cal D} \phi \ \bar \Psi [ \phi ] \phi({\bf x}_1) \dots \phi({\bf x}_n) \Psi [ \phi ] 
\nonumber \\
&&= {1\over Z} \int {\cal D} \phi \ \phi({\bf x}_1) \dots \phi({\bf x}_n) \ {\rm e}^{-\lambda  \int d^2 {\bf x} \ \partial_i \phi \partial_i \phi} 
\label{two}
\end{eqnarray}
with $Z=\int {\cal D} \phi \ \bar \Psi [ \phi ] \Psi [ \phi ]$. These are nothing else than the correlations of the Gaussian model \cite{difra}. This $c=1$ universality class of quantum critical behavior has been called quantum Lifshitz theory \cite{fra}, since the prototype (non-relativistic) action leading to the ground state functional (\ref{one}) describes also three-dimensional Lifshitz points \cite{lif}. Several examples of theoretical models in this quantum universality class have been described in \cite{fra}. 

In this paper we shall describe another $c=1$ conformal quantum critical point, one that is realized in nature as the superconducting-insulating (SI) quantum phase transition in 2D Josephson junction arrays. In addition we argue that the SI quantum transitions in magnetically frustrated arrays realize other universality classes of quantum critical behavior, those corresponding to the unitary minimal models \cite{difra} with central charge $c=1-6/m(m+1)$.

JJAs are quadratic, planar arrays of superconducting islands with nearest neighbours Josephson couplings of strength $E_J$. Each island has a capacitance $C_0$ to the ground;
moreover there are also nearest neighbours capacitances $C$. The Hamiltonian characterizing such systems is thus given by
\begin{eqnarray}H &&= \sum _{<{\bf x \bf y}>}
\left( {C\over 2} \left( V_{\bf y}-V_{\bf x} \right) ^2   
+ E_J \left( 1-{\rm cos}\ N\left( \Phi _{\bf y} - \Phi _{\bf x} \right) \right) \right) 
\nonumber \\
&&+ \sum_{\bf x} \ {C_0\over 2} V_{\bf x} 
\label{hJJa}
\end{eqnarray}
where boldface characters denote the sites of the two-dimensional array, $<{\bf x \bf y}>$ indicates nearest neighbours, $V_{\bf x}$ is the electric potential of the island at ${\bf x}$ and $\Phi _{\bf x}$ the phase of its order parameter.  The physics of JJAs is governed by the competition between quantized charges $Q = q_e j^0 = q_e N n$, $n\in \mathbb{ Z}$ determining the potential via the Coulomb law $\left( C_0 - C \Delta \right) V_{\bf x} = q_e j^0_{\bf x}$ and their canonically conjugate quantized vortices $\phi = \phi^0 / q_e = n 2\pi /Nq_e$, $n\in \mathbb{ Z}$, representing topological defects in the superconducting phase configuration. For generality we allow for any integer $N$ in the Josephson coupling, so that the phase has periodicity $2\pi /N$: obviously $N=2$ (Cooper pairs) for the real systems.

In this paper we shall focus on the nearest-neighbours capacitance limit $C\gg C_0$, which is accessible experimentally and we will thus henceforth set $C_0=0$. In this regime the charging energy $E_C = q_e^2/2C$ and the Josephson coupling $E_J$ are the two relevant energy scales in the problem. These can be traded for one mass scale $\sqrt{2N^2E_CE_J}$ representing the Josephson plasma frequency and one massless parameter $E_C/E_J$ which governs the quantum phase structure. 

In \cite{dst1} we have shown that planar JJAs can be mapped onto the strong coupling limit of a (2+1)-dimensional Abelian gauge theory with a mixed Chern-Simons (CS) term \cite{djt} (we use Greek letters for Minkowski space-time indices and Latin letters for 2D space indices), 
\begin{equation}{\cal L} = - {1\over 4e^2}f_{0i} f^{0i} + {k \over 2\pi }
a_{\mu } \epsilon^{\mu \alpha \nu } \partial_{\alpha } b_{\nu } - {1\over 4g^2}
g_{0i }g^{0i } \ .
\label{mcss}
\end{equation}
where $f_{\mu \nu} = ( \partial_{\mu} a_{\nu}- \partial_{\nu} a_{\mu})$, $g_{\mu \nu} = ( \partial_{\mu} b_{\nu}- \partial_{\nu} b_{\mu})$. 

The idea of this mapping is that $j^\mu = ({k^{3/2}/2\pi}) \epsilon^{\mu \nu \alpha} \partial_\nu b_\alpha$ represents the conserved current of charges while $\phi^\mu = ({1/2\pi k^{1/2}}) \epsilon^{\mu \nu \alpha} \partial_\nu a_\alpha$ represents  the conserved current of vortices. The mixed Chern-Simons coupling generates the Aharonov-Bohm interactions between charges and vortices. The "electric" field $g^{0i}$ encodes both the Josephson currents (transverse part) and the Coulomb interaction between vortices (longitudinal part). Correspondingly, the "electric" field $f^{0i}$ encodes the Coulomb interactions between charges (longitudinal part). The transverse part of $f^{0i}$ represents a bare kinetic term for the vortices: this contribution, which is absent in the original Hamiltonian (\ref{hJJa}), has to be added in order to expose the full duality between charges and vortices. Note, however, that such a vortex kinetic term is anyhow radiatively induced. This near-duality between charges and vortices has been often invoked in the literature \cite{jos} to explain the experimental quantum phase diagram
at very low temperatures. The mapping (\ref{mcss}) becomes exact in this self-dual approximation. This approximation may lead to a small displacement of the exact critical point but should not affect universal quantities. 

The two massive parameters $e^2$ and $g^2$ are directly related to the JJA scales: $e^2 = 2NE_C$, $g^2 = (4\pi ^2/N) E_J$, while the dimensionless Chern-Simons coupling $k$ determines the charge of the condensate $k=N$. The quantization of charges and vortices requires the gauge symmetries to be compact $U(1)$ symmetries. On the lattice, the Chern-Simons coupling between the two degrees of freedom is thus invariant under shifts of the gauge link variables \cite{dst1}
\begin{equation}
a_{\mu } \to a_{\mu } + {2\pi \over \sqrt k}  n_{\mu } \ ,  b_{\mu } \to b_{\mu } + {2\pi \over \sqrt k} \ m_{\mu } \ ;  n_{\mu },m_{\mu } \in \mathbb{ Z}\ ,
\label{sh}
\end{equation}
so that $j^0 = k n$, $n\in \mathbb{ Z}$ and $\phi^0 = n 2\pi /k$, $n\in \mathbb{ Z}$. 

The compactness of the gauge fields implies the existence of topological defects in the mixed Chern-Simons model. These represent the quantized charges and vortices of JJAs, the propagating modes with topological mass \cite{djt} $m_{\rm top}=eg/2\pi$ corresponding instead to Josephson plasma oscillations. The condensation (or lack thereof) of these topological defects determines the phase structure of the theory. When electric topological defects (charges) condense, the array is a global superconductor, when magnetic topological defects (vortices) condense, the arrays is an insulator. Since the model is self-dual, the SI quantum phase transition at $T=0$ lies exactly at the self-dual point $e/g =1$ \cite{dst1}. This quantum critical point is characterized by the fact that both topological excitations are dilute and, thus, is characterized by the continuum model (\ref{mcss}) with $e=g$. 

This model (\ref{mcss}) is a strongly coupled gauge theory in the sense that the magnetic permittivity  is much larger than the electric permeability, so that magnetic fields become negligible and the corresponding magnetic terms are absent from the action. This form of non-relativistic strong coupling limit of Maxwell-Chern-Simons theory has already been addressed in \cite{ggp} for the case of a single gauge field. 

To show that the quantum critical point is a conformal one we proceed to canonically quantize the theory (\ref{mcss}) at this point, following \cite{djt}. The quantization will be performed in the Weyl gauge $a^0=0$  and  $b^0 = 0$. The two canonical momenta conjugate to the canonical variables $a^i$ and $b^i$ are:
\begin{eqnarray} \Pi^i_a &=& {\delta {\cal L} \over  \delta  (\partial_0 a^i )} = - {1\over e^2}f^{0i} +  {k \over 2\pi }  \epsilon^{ij} b^j \nonumber \\
\Pi^i_b &=& {\delta {\cal L} \over  \delta  (\partial_0 b^i )} = - {1\over g^2}g^{0i} +  {k \over 2\pi }  \epsilon^{ij} a^j \ . 
\label{mc}
\end{eqnarray}
The Hamiltonian density, when written in canonical variables takes the form
\begin{equation}H = {e^2 \over 2} \left(\Pi^i_a -  {k \over 4\pi } \epsilon^{ij} b^j \right)^2 +  {g^2 \over 2} \left( \Pi^i_b -  {k \over 4\pi } \epsilon^{ij} a^j \right)^2 \ .
\label{ham}
\end{equation}
The Gauss laws, implementing gauge invariance, are given by:
\begin{eqnarray} G_a \equiv \partial_i\Pi^i_a &+&  {k \over 4\pi } \partial_i \epsilon^{ij} b^j 
\approx 0 \nonumber \\
G_b \equiv \partial_i\Pi^i_b &+&  {k \over 4\pi } \partial_i \epsilon^{ij} a^j 
\approx 0\ , 
\label{gl}
\end{eqnarray}
and must be imposed as  conditions on physical states: 
\begin{equation} G_a  \Psi [a^i,b^i]   = G_b  \Psi [a^i,b^i]  = 0  \ .
\label{glc}
\end{equation}
In this functional Schr\"odinger representation the energy eigenstates satisfy:
\begin{equation} H   \Psi [a^i,b^i]  = E  \Psi [a^i,b^i] \ ,
\label{eat}
\end{equation}
with the subsidiary conditions( \ref{glc}). Following \cite{djt} we write the ground state functional as the product of a cocycle and a part that depends only on the transverse part of the two dynamical variables, $a^i_T$ and $b^i_T$: 
\begin{eqnarray}&\Psi_0 [a^i,b^i] =   \exp  i \chi  [a^i,b^i]  \ \Phi (a^i_T,b^i_T) \ , \nonumber \\
& \chi [a^i,b^i]  = {k \over 4 \pi} \int d^2x \left[ b {\partial_i \over \Delta  } a^i   + a {\partial_i \over \Delta  } b^i \right]  \ , \nonumber \\
 & \Phi [a^i_T,b^i_T] = \exp   {-  k \over 4 \pi} \int d^2x \left( {g \over e} ( a^i_T )^2 + {e \over g} ( b^i_T  )^2  \right) \ , 
\label{gswf}
\end{eqnarray}
where  $a= \epsilon^{ij} \partial_i a^j ; b = \epsilon^{ij} \partial_i b^j$, $\Delta = \partial_i \partial_i$ and 
$$a^i_T  =  P^{ij}  a^j (x)  \ ,\  b^i_T(x) =  P^{ij}  b^j \  ,$$
with the projector $ P^{ij} $ onto the transverse part of the gauge fields given by: $ P^{ij} = \left (\delta^{ij} -  {\partial^i  \partial^j\over \Delta  } \right) $. 
The ground state energy vanishes  after subtracting the (infinite) zero point energy $E_0 =  {k \over  \pi} {\rm Tr} P$.
Using the Hodge decomposition for the spatial components of the two gauge fields $a^i$ and $b^i$:
\begin{eqnarray}a_i &=& \partial_i \xi + \epsilon_{ij} \partial_j \phi \ , \nonumber \\
b_i &=& \partial_i \lambda + \epsilon_{ij} \partial_j \psi \ ,
\label{hod}
\end{eqnarray}
we can rewrite $\Psi_0 [a^i,b^i]$ as:
\begin{eqnarray}&\Psi_0 [a^i,b^i]  = \exp  {- i k \over 4 \pi} \int d^2x \left[ \psi \Delta \xi + \phi \Delta \lambda \right] \times \nonumber \\
&\times \exp  { - k \over 4 \pi} \int d^2x \left[   {g \over e} (\partial_i \phi)^2 +  {e \over g} (\partial_i \psi)^2\right] \ .
\label{ngs}
\end{eqnarray}

Two are the important points to stress about this result. First of all, the imaginary cocycle does not enter the computation of probabilities and equal time correlation functions. The longitudinal variables $\xi$ and $\lambda$ are thus irrelevant for these quantities, which is a consequence of gauge invariance. Secondly, in the strong coupling limit, the familiar non-local kernel for the transverse gauge fields in the ground state wave functional of Maxwell-Chern-Simons theory \cite{djt} becomes local, a simple contact term for $a^i_T$. This is the main origin of the appearance of the two free bosons in the ground state wave functional above. 

As a consequence of this, equal-time expectation values of local operators ${\cal O}_i$ of the dynamical fields $\phi$ and $\psi$ in the ground state take the form:
\begin{eqnarray}&\langle {\cal O}_1[\phi (x_1)]  .......  {\cal O}_n[\phi (x_n)]   {\cal O}_1[\psi (x_1)]  ......  {\cal O}_m[\psi (x_m)]  \rangle = \nonumber \\
&=  {1 \over Z}  \int   {\cal D} \phi {\cal D} \psi   {\cal O}_1[\phi (x_1)]  ..  {\cal O}_n[\phi (x_n)]   {\cal O}_1[\psi (x_1)]  ..  {\cal O}_m[\psi (x_m)]   \nonumber \\
&\times    \exp   {- k \over 2 \pi} \int d^2x \left[ {g \over e}  (\partial_i \phi)^2 +   {e \over g} (\partial_i \psi)^2\right]  \ ,
\label{prob}
\end{eqnarray}
with the normalization $Z$ given by $$Z=\int {\cal D} \phi {\cal D} \psi \ \bar \Psi_0 [ \phi, \psi] \Psi_0 [ \phi , \psi] \ .$$

Equation (\ref{prob}) implies that the equal-time quantum correlations of the dynamical fields $\phi$  and $\psi$  are nothing else than the correlations of the (doubled) Gaussian model \cite{difra} with central charge $c=1$. The universality class of the quantum SI critical point  is thus that of quantum Lifshitz theory \cite{fra}. 

Josephson junction arrays show a most interesting behavior in presence of offset charges on the superconducting islands or when subject to a uniform external magnetic field $B$ \cite{bla}. In this latter case, the role of the dimensionless "magnetic charge" assigned to the islands is the frustration parameter $f = {\Phi \over \Phi_0}$, where $\Phi = BA$ is the magnetic flux piercing the array of area $A$ and $\Phi_0 = 2\pi /\kappa$ is the quantum of magnetic flux. In the self-dual approximation, the two effects are the exact mirror of each other; we shall thus discuss only the latter case of magnetic frustration and concentrate exclusively on the correlations of the field $\psi$. 

What has been observed in experiments \cite{zant} is that, for particular fractional values of the frustration $f$, the array behaves in a similar way as with zero magnetic field with the only difference that the position of the SI quantum phase transition is displaced from the self-dual point $E_C/E_J= \pi^2 /2$ to lower values. In other words, at specific values of the frustration $f$, the quantum critical point is conformal even in presence of an external dimensionful perturbation. 

In particular, for square arrays, in which the frustration is due exclusively to the magnetic field, the observed fractions are $f = 1/2, 2/5, 1/3, 1/4,...$. For fully frustrated arrays with $f=1/2$, the critical parameter $E_C/E_J$ for the quantum transition is lowered by approximately 0.7 with respect to the non-frustrated case, with measurements having an error margin of 10\%. We are not aware of reliable measurements of the critical point for other frustration fractions. 

These results are interpreted as the formation of a vortex lattice commensurate with the underlying junction network, which completely shields the external magnetic field. A scaling theory of the 2D SI quantum phase transition based on the Bose condensation of these vortices has been developed by Fisher \cite{fis}. Here we would like to stress that this mechanism is very reminiscent of the binding of an even number of magnetic fluxes by electrons forming quantum Hall fluids at fractional filling \cite{joh}, with the only difference that in the present situation the magnetic field is totally screened. In the quantum Hall framework it is well known that the emerging composite electrons can be described by dressed vertex operators of conformal field theories \cite{mus}. We thus suggest to interpret the experiments on frustrated Josephson junction arrays in this light. 

To make contact with the standard conformal field theory notation we shall proceed to a rescaling:
\begin{equation} \phi \rightarrow \sqrt{ k } \phi  \ ,  \psi \rightarrow   \sqrt{ k } \psi \ .
\label{res}
\end{equation}
This has the consequence that the fields $\phi$ and $\psi$ originally compactified on a circle of radius $2\pi /\sqrt k$( see eq.  (\ref{sh})) acquire the periodicity
\begin{equation} \phi \equiv  \phi + 2 \pi  \ , \ \psi \equiv  \psi + 2 \pi  \ ,
\label{sp}
\end{equation}
and the ground-state functional becomes
\begin{equation}\Psi_0 [ \phi , \psi]  =  \exp {- 1 \over 4 \pi} \int d^2x \left[ {g \over e}  (\partial_i \phi)^2 +   {e \over g} (\partial_i \psi)^2\right] \ ,
\label{nprob}
\end{equation}
where we have dropped the cocycle that does not contribute anyhow to probabilities.

We shall thus assume that the quantum critical point in presence of magnetic frustration is described by exactly the same quantum ground state wave functional (\ref{nprob}) but that a background charge representing the contribution of the shielding vortex lattice has to be included in vacuum expectation values. In conformal field theory \cite{difra} this is known as the Feigin Fuchs representation \cite{zub}, in which the shielding is formally implemented by a charge at infinity: 
\begin{eqnarray}
&&\langle {\rm e}^{i \sqrt{2} \alpha_1\psi({\bf x}_1)} \dots {\rm e}^{i \sqrt{2} \alpha_n \psi({\bf x}_n)} \rangle _f = {\rm lim}_{R\to \infty}
\nonumber \\ 
&& R^{f^2g/e}
\langle {\rm e}^{i \sqrt{2} \alpha_1\psi({\bf x}_1)} \dots {\rm e}^{i \sqrt{2} \alpha_n \psi({\bf x}_n)}  {\rm e}^{-i \sqrt{2} f \psi(R)} \rangle \ ,
\label{tone}
\end{eqnarray}
where the second expectation value is computed with the original ground state wave functional (\ref{nprob}). 
When the external magnetic field is completely shielded, the charge at infinity is given exactly by the full frustration, so that the only non-vanishing correlations have $\sum_i \alpha^i = f$. This corresponds to a shift of the central charge and of the conformal dimensions \cite{zub}
\begin{eqnarray}
 c=1 && \to  \  c= 1- 6{f^2g\over4e} \ , 
 \nonumber \\
 h_i = {\alpha^i g\over 4 e} &&\to \  h_i = {g \alpha^i (\alpha^i -f) \over 4e} \ .
 \label{ttwo}
 \end{eqnarray}
 
For particular values of the frustration $f$ and of the coupling $e/g$ there exist conformal field theories with a finite number of primary fields, for which the bootstrap algebra closes: these theories are called "minimal models" since they have a minimal number of possible excitations with respect to generic models for any values of $f$ and $e/g$ and are thus particularly robust. It is to be expected that the observed quantum conformal critical points in presence of external magnetic fields correspond exactly to these minimal models. This is the same stability principle that led to the prediction of the Jain hierarchy of quantum Hall fractions \cite{joh} from the minimal models of the infinite-dimensional algebra of quantum area-preserving diffeomorphisms \cite{ctz}. 

The minimal models exist for $f=2(p-q)/q$ and $e/g=p/q$ for $p$ and $q$ co-prime integers. However, only if $|p-q|=1$ are the corresponding theories unitary. Unitary minimal models can be parametrized by a single integer $m \ge 3$ \cite{zub} and we obtain thus the following series of models
\begin{equation}
f = {2 \over m+1} \ , \ {e\over g} = {m\over m+1} \ , \ m \ge 3 \ .
\label{tthree}
\end{equation}
The first few predicted frustrations are thus
\begin{equation}
f=1/2, 2/5, 1/3, 2/7, 1/4 \dots \ .
\label{tfor}
\end{equation} 
Apart from 2/7, these are exactly the observed frustrations where a quantum critical point similar to the $f=0$ one exists. Note also that the predicted critical coupling for the fully frustrated model with $f=1/2$ is $e/g=3/4=0.75$ well within the 10\% error margin of the experimental value 0.7. 

\noindent {\bf Acknowledgement} MCD thanks CERN for hospitality  during the period in which this work was completed.

\end{document}